\documentclass[prb,twocolumn,showpacs]{revtex4}
\newcommand{\BF}[1]{\mbox{\boldmath $#1$}}
\def\mn#1{*\marginpar[\scriptsize*#1]{\scriptsize*#1}}

\usepackage{graphicx}
\usepackage{dcolumn}
\usepackage{bm}

\begin{document}

\title{Two different scaling regimes in Ginzburg-Landau model with
Chern-Simons term}
\author{Hagen Kleinert}
\email{kleinert@physik.fu-berlin.de}
\homepage{http://www.physik.fu-berlin.de/~kleinert/}
\author{Flavio S. Nogueira}
\email{nogueira@physik.fu-berlin.de}
\affiliation{Institut f\"ur Theoretische Physik,
Freie Universit\"at Berlin, Arnimallee 14, D-14195 Berlin, Germany}

\date{Received \today}

\begin{abstract}
The Ginzburg-Landau model
with a Chern-Simons term is shown to
possess
two different scaling regimes depending
on whether the mass
of the scalar field is zero or not.
In contrast to  pure $\phi^4$ theories,
the
Ginzburg-Landau model
with a
topologically
generated mass
exhibits
quite different properties
in perturbation theory. Our analysis
suggests that the two scalings 
could coincide at
a non-perturbative level. This view
is supported by a $1/N$-expansion in the  massive scalar field
regime.
\end{abstract}

\maketitle

\section{Introduction}

The Ginzburg-Landau (GL) model\cite{cGL} with the Lagrange density
\begin{eqnarray}
{\cal L}_{\rm GL}=\frac{1}{2}(\BF{{\nabla}}\times{\bf a})^2
+|(\BF{\nabla}-iq{
\bf a})\phi|^2
+r|\phi|^2+\frac{u}{2}|\phi|^4
\label{@GL}\end{eqnarray}
was
set up more than 50 years ago
to describe
 superconductivity. Since then it has
 been used to describe
 a variety of other
physical systems\cite{KleinertBook},
 where complex order field $\psi({\bf x})$ and vector potential
${\bf a}({\bf x})$
are not related to Cooper pairs and magnetic fields.
 An important application deals with
smectic liquid crystals
where $\psi({\bf x})$ describes the smectic order
and
${\bf A}({\bf x})$
the transverse displacement
of the nematic director
 \cite{deGennes}.
Other fascinating applications
arise by adding a
topological Chern-Simons (CS) term \cite{Deser} for the vector potential
to the GL model:
\begin{eqnarray}
{\cal L}_{\rm CS}&=&
\frac{i\theta}{2}{\bf a}\cdot(\BF{\nabla}\times{\bf a}).
\label{@CS}\end{eqnarray}
In this case one speaks of
a
 Chern-Simons-Ginzburg-Landau (CSGL) model.
This model possesses properties
found in
the famous fractional quantum Hall effect, where
the coupling parameter of the CS term
determines
a nontrivial
 phase factor
for the exchange of two complex fields
and thus
 the statistics of Laughlin quasi-particles
\cite{GirvinBook}. Without an initial
Maxwell term $(\BF{\nabla}\times {\bf a})^2/2$,
this interpretation
as been advanced
by Zhang\cite{Zhang}.
All gradient terms in his effective action
are caused  by
fluctuations
of a vector potential
with only a CS term.
But also
the theory with a Maxwell term has physical
significance
 since
it emerges naturally by
when constructing a dual
disorder model of CSGL model without  the Maxwell term
\cite{KleinNog}.
In addition, such a model has been
found
when bosonizing
theories of strongly interacting fermions
in three dimensions \cite{Nayak,Fradkin}.

Another interesting application of the CSGL model
without a Maxwell term
arises
in the
field theoretical approach to polymers,
where the degree of entanglement
is controlled by the parameter $\theta$\cite{PI}.
Detailed results have
 been obtained  recently \cite{FerrKleinLazz,FerrLazz}.

In this paper we shall discuss the fixed point structure of the
the CSGL model with a Maxwell term.
The Lagrangian of the model is
\begin{eqnarray}
\label{Lspin}
{\cal L}_{\rm CSGL}
={\cal L}_{\rm GL}
+{\cal L}_{\rm CS}.
\end{eqnarray}
The fixed-point structure of a standard GL
Lagrangian has been investigated at various places\cite{Herbut,Folk,KleinNog1}.
In the presence
of a
CS term,
it has been discussed in Refs. \onlinecite{KleinertCS}
and \onlinecite{deCalan}. It must be emphasized
that in these references
{\it the Maxwell term is explicitly included}
in contrast to earlier work
where it was ignored
 \cite{Semenoff}.
This makes an important differences in the fixed-point
structure since in the  presence Maxwell term,
 the charge
is  no longer dimensionless, and there is
the generation of
another mass called the
 topological mass.
An important result of
Ref. \onlinecite{deCalan} was that, although the CS term is not renormalized,
the $\beta$-function of the topological coupling {\it is not} zero
due to the presence of the Maxwell term.

It was noted by Semenoff \cite{Semenoff} that the renormalization
of the CSCL model depends on
the mass of the  scalar field. Kleinert and Schakel
\cite{KleinertCS}
considered the CSGL  and derived scaling laws as a function
of
the
renormalized mass of the scalar field. Later, de Calan
{\it et al.} \cite{deCalan} considered the same model, but
within
 renormalization group (RG) approach at
the
critical point. They obtained
considerably more involved
RG functions than those of
Ref. \onlinecite{KleinertCS}.

In this paper we shall
improve considerably the discussions of Refs.
\onlinecite{KleinertCS} and \onlinecite{deCalan} and
exhibit the relation between both scaling behaviors at the one-loop level.
The plan of the paper is the following. In Section II we discuss an effective
mean-field theory which only includes fluctuations of the gauge field.
{}From this we can already observe a
particular
feature coming from the CS term: for  large $\theta$, there is
no tricritical point and therefore no first-order phase transition,
in contrast to the pure GL model\cite{Kleinert}.
This approximation, however, does not
 distinguish
reliably
critical from tricritical behavior.
For this we
employ in Section III the
RG to obtain information
on the phase transition. At the one-loop level
 we are then
able
to
distinguish {\it two different scaling behaviors with quite different
physical properties}.
The relation between them
is illuminated
in Section IV
by comparing a non-perturbative $1/N$ expansion in the nonzero-mass regime
 with
the one-loop approximation in the massless
regime.
The
qualitative behavior of the  non-perturbative result
 shows a remarkable agreement with
the one-loop approximation.
In Section V we obtain the exponents for the ``fermionic'' fixed point,
which is reached for a specific value of the CS coupling parameter
obtained
from the bosonization scheme \cite{Fradkin,Zhang}.
A final  discussion is given
 in Section VI.

\section{Effective mean-field theory}

Let us integrate out the
vector potential in the effective action
to derive
a lowest-order
effective
 mean-field
theory for the model (\ref{Lspin}), by analogy
with the
procedure
on the pure GL theory
by Halperin {\it et al.}
\cite{HLM}. For a uniform order
field
$\phi\equiv\phi_0/\sqrt{2}$, where $\phi_0$ is a real
constant,
this operation can be done exactly.
The result is a a free energy density
\begin{eqnarray}
\label{HLMCS} &&\!\!\!\!\!\!\!\!
{\cal F}\!\!=-\frac{1}{12\pi}\{[M_+^2(\phi_0^2)]^{3/2}
\!+[M_-^2(\phi_0^2)]^{3/2}\}
+\frac{r}{2}\phi_0^2
+\frac{u}{8}\phi_0^4,
\nonumber \\&&
\end{eqnarray}
where
\begin{equation}
M_\pm^2(\phi_0^2)\equiv q^2\phi_0^2+
\frac{\theta^2}{2}\pm\frac{|\theta|}{2}\sqrt{\theta^2+
4q^2\phi_0^2}.
\end{equation}
In (\ref{HLMCS})
dropped field-independent infinite
term
 and absorbed a term proportional to the
ultraviolet cutoff in $r$.
For $\theta=0$, our result
reduces to the usual Halperin-Lubensky-Ma
(HLM) expression\cite{HLM}
which
displays a first-order phase transition.
This remains true
for
sufficiently small $\theta\neq0$.
At larger
 values
of $\theta$, however,
 the transition
is of second-order.
This change of order
is quite subtle:
if we
expand ${\cal L}^{\rm eff}$ \`a la Landau up to
the power $\phi_0^4$, we obtain for a constant order field
\begin{equation}
\label{expHLMCS}
{\cal F}_{\rm L}\simeq-\frac{|\theta|^3}{12\pi}+\left(\frac{r}{2}
-\frac{q^2|\theta|}{4\pi}\right)\phi_0^2
+\frac{u}{8}\phi_0^4,
\end{equation}
and we see that the above equation does not have the correct
$\theta\to 0$ limit, being valid only for large
$\theta$. The free energy  (\ref{expHLMCS}) has only
a second-order phase transition, with a
 critical point
at $r_{c}=q^2|\theta|/2\pi$. Note that
in contrast to the GL case
there is no cubic term in $\phi_0$,
and that the $\phi_0^4$-term receives
no contribution from
the CS coupling---both properties
would have generated a
tricritical point in this
approximation.
The latter property implies
that the one-loop gauge field
graph with four external scalar field lines vanishes
if the external momenta are set to zero,
as a peculiar
feature
of the CSGL model
noted earlier in Ref. \onlinecite{KleinertCS}. The
same graph is, however, non-vanishing
at nonzero external momenta.
 This is the origin of
the
 two
different scaling regimes
in this model which we want to dicuss in this paper.

\section{Renormalization group functions}

We now calculate the RG functions of the   model.
 As discussed in Section II, the scaling with a finite
scalar field mass looks different from the one where such a
mass is absent, as
noted by
Semenoff
\cite{Semenoff}. Let us study these scaling behaviors separately and
see how the are related to each other.
 Many of the results of this
section have been obtained
before in Refs.
\onlinecite{KleinertCS,deCalan}. However, a discussion on the relation
between the tow scaling behaviors
of Ref. \onlinecite{KleinertCS} and
 Ref. \onlinecite{deCalan} is new.
This relation will require an extension
of the CSGL model
 (\ref{Lspin})
to $N/2$ complex scalar field.

\subsection{Massive Scaling Regime}

In the scaling regime with a massive
scalar field the
propagators are given by
\begin{equation}
\label{G}
G({\bf p})=\frac{1}{{\bf p}^2+r},~~~r\neq0.
\end{equation}
for the scalar and
\begin{equation}
\label{D}
D_{\mu\nu}({\bf p})=\frac{1}{{\bf p}^2+\theta^2}\left(\delta_{\mu\nu}
-\frac{p_\mu p_\nu}{{\bf p}^2}-\theta\epsilon_{\mu\nu\lambda}
\frac{p_\lambda}{{\bf p}^2}\right)
\end{equation}
for the vector field in
the Landau gauge.
The Lagrangian is written in terms of renormalized quantities as

\begin{widetext}
\begin{equation}
L=\frac{Z_a}{2}(\BF{\nabla}\times{\bf a}_r)^2+
\frac{i\theta_r}{2}{\bf a}_r\cdot(\BF{\nabla}\times{\bf a}_r)
+Z_\phi|(\BF{\nabla}-iq_r{\bf a}_r)\phi_r|^2+
Z_\phi^{(2)}m^2|\phi_r|^2+\frac{Z_g m g}{2}|\phi_r|^4.
\end{equation}
\end{widetext}
The renormalized fields are given by $\phi_r=Z_\phi^{-1/2}\phi$ and
${\bf a}_r=Z_a^{-1/2}{\bf a}$, and  we have set
$u_r\equiv m g$ to have a
 dimensionless coupling constant
 $g$.
We also have introduced a mass of the scalar field $m$ by
$m^2\equiv {Z_\phi^{(2)}}^{-1}Z_\phi r$.
Note  that the CS term is
not renormalized \cite{Semenoff}, implying that
$\theta_r=Z_a\theta$. The renormalized charge is $q_r=Z_a^{1/2}q$.
We introduce two
 dimensionless gauge coupling constants by
$t\equiv \theta_r/m$ and $f\equiv q_r^2/m$.
The renormalization constants are fixed by
imposing
normalization
conditions for the one-particle irreducible two- and four-point
functions:
\begin{equation}
\label{cond1}
\Gamma_{r,11}^{(2)}(0)=m^2,
\end{equation}
\begin{equation}
\label{cond2}
\left.\frac{\partial\Gamma_{r,11}^{(2)}}{\partial{\bf p}^2}
\right|_{{\bf p}=0}=1,
\end{equation}
\begin{equation}
\label{cond3}
\Gamma_{r,1111}^{(4)}(0,0,0,0)=3mg,
\end{equation}
\begin{equation}
\label{cond4}
\Gamma_{r,11}^{(1,2)}(0,0,0)=1,
\end{equation}
\begin{equation}
\label{cond5}
\left.\frac{\Gamma_{r,\mu\mu}}{\partial{\bf p}^2}\right|_{{\bf p}=0}=2.
\end{equation}
Let us define the RG functions:
\begin{widetext}
\begin{equation}
\label{gammaphi}
\!\!\!\gamma_\phi\equiv m\frac{\partial\ln Z_\phi}{\partial m},\,
\label{gammaa}
~~~~~~\gamma_a\equiv m\frac{\partial\ln Z_a}{\partial m},  \,
\label{gamma2}
~~~~~~\gamma_\phi^{(2)}\equiv m\frac{\partial\ln Z_\phi^{(2)}}{\partial m}.
\end{equation}
Within the present renormalization scheme, these functions are
given explicitly in the one-loop approximation by
\begin{equation}
\label{gammaphi1}
\!\!\!\!\gamma_\phi=-\frac{2}{3\pi}\frac{f}{(1+|t|)^2},\,
\label{gammaa1}
~~~~~~\gamma_a=\frac{Nf}{48\pi}, \,
\label{gamma21}
~~~~~~\gamma_\phi^{(2)}=-\frac{(N+2)g}{16\pi}.
\end{equation}
\end{widetext}
The $\beta$-functions are given by
\begin{equation}
\label{betaf}
\beta_f\equiv m\frac{\partial f}{\partial m}=(\gamma_a-1)f,
\end{equation}
\begin{equation}
\label{betat}
\beta_t\equiv m\frac{\partial t}{\partial m}=(\gamma_a-1)t,
\end{equation}
\begin{equation}
\label{betag}
\beta_g\equiv m\frac{\partial g}{\partial m}=(2\gamma_\phi-1)g
+\frac{N+8}{16\pi}g^2.
\end{equation}
Note the absence of a
term proportional to $f^2$ in Eq.~(\ref{betag}). This
generalizes
the observation
in the previous approximation
that the
$\phi_0^4$-term in the Landau
 expansion (\ref{expHLMCS}) is $\theta$-independent.
 This is in contrast
to the
pure GL model where a $f^2$-term is present in $\beta_g$
\cite{HLM,Lawrie,Uzunov,Folk,Herbut}.
The present absence of the $f^2$-term is the reason for
 the
existence of a charged fixed point
(which remains true for all values of $N$, if the model is extended to $N/2$
complex fields,
in contract to the pure GL model where $N>365$ is needed
as pointed out by
 Halperin {\it et al.} \cite{HLM}.

The anomalous dimension
of the gauge field    $\eta_a\equiv\gamma_a^*$
has an interesting property. From Eq.~(\ref{betaf})
we see that
$\eta_a=1$,
 implying a fixed point also in
(\ref{betat}) for any $t$, which means
that the critical exponents can vary  continuously. The charged fixed point is given by
\begin{equation}
f_*=48\pi/N,~~~\label{g*}
g_*=\frac{16\pi}{N+8}\left[\frac{64}{N(1+|t_*|)^2}+1\right].
\end{equation}
Note
that $\beta_t$ does not vanish identically
as in the absence
of
a
Maxwell term. It does, however, vanish 
at the fixed point where $ \gamma _a=1$
 for all values of $t$,
which has the same effect as
$\beta_t\equiv 0$, thus allowing
 for arbitrary
fixed-point values $t^*$.

\begin{figure}
\begin{picture}(87.87,143.05)
\unitlength.7mm
\put(-36,3){\includegraphics{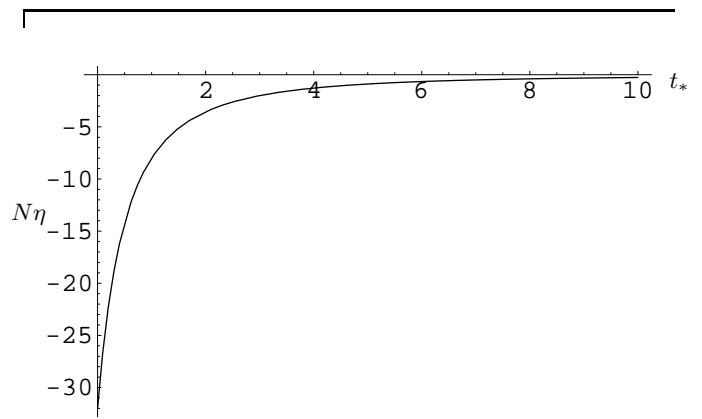}}
\put(83,68){$t_*$}
\put(-42,43){$N\eta$}
\end{picture}
\caption{\label{Figeta1}Plot of $N\eta$, where $\eta$ is given
by Eq. (\ref{eta}), as a function of $t_*>0$.}
\end{figure}

\begin{figure}
\begin{picture}(87.7,143.05)
\unitlength.7mm
\put(-36,3){\includegraphics{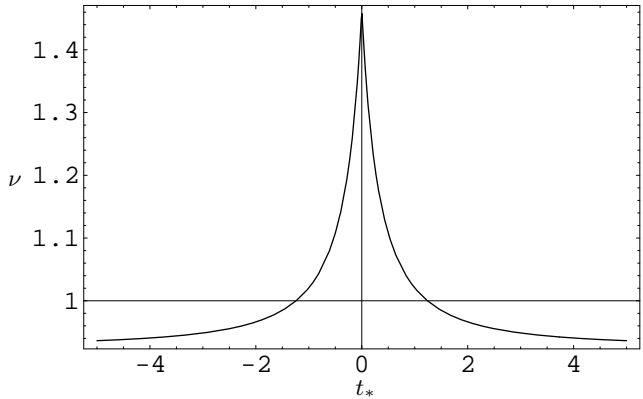}}
\put(26,0.5){$t_*$}
\put(-40,40){$\nu$}
\end{picture}
\caption{\label{Fignu1}Plot of the critical exponent $\nu$ as a function of
$t_*$ for $N=68$.}
\end{figure}

The critical exponent $\eta$ is given by the fixed point value of the
RG function $\gamma_\phi$:
\begin{equation}
\label{eta}
\eta=-\frac{32}{N(1+|t_*|)^2}.
\end{equation}
Thus, although a charged fixed
point exists for all values of $N$,
the fixed points with $N\leq 32/[(1+|t_*|)^2]$
are unphysical since the inequality $\eta>-1$ is not fulfilled.
Thus, we still have a critical value of $N$, but considerably
smaller than the value $N_c=365$ of Ref.~\onlinecite{HLM}.
For any fixed $N\leq N_c=32$,
there are
physical critical exponents
provided
that
\begin{equation}
\label{tineq}
|t_*|>t_c(N)=4\sqrt{\frac{2}{N}}-1.
\end{equation}
In Fig. \ref{Figeta1} we plot $N\eta$ as a function of $t_*$
for
positive values of $t_*$.
The critical exponent $\nu$ is obtained as the fixed-point value of
the RG function
$ 
\label{nuphi}
\nu_\phi={1}/[2+\gamma_\phi^{(2)}-\gamma_\phi],
$
which is
\begin{equation}
\label{nu}
\nu=\left[2-\frac{N+2}{N+8}-\frac{N-4}{N+8}
\frac{32/N}{(1+|t_*|)^2}\right]^{-1}.
\end{equation}
This is plotted in Fig.~\ref{Fignu1}  as a function of
$t_*$ for $N=68$.

\subsection{Massless Scaling Regime}

\begin{figure}
\begin{picture}(87.7,143.05)
\unitlength.7mm
\put(-36,3){\includegraphics{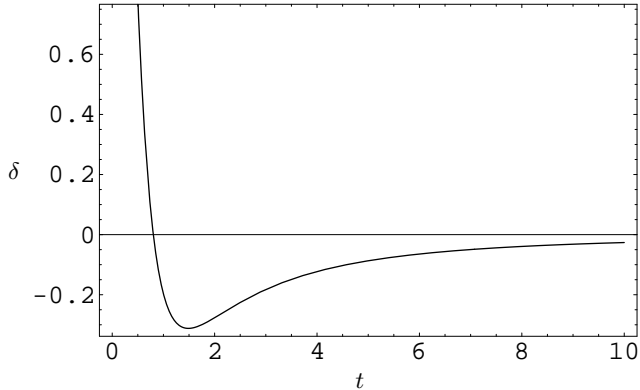}}
\put(26,0.5){$t$}
\put(-40,40){$\delta$}
\end{picture}
\caption{\label{Figomega}Plot of $\delta$ as a function of $t$.}
\end{figure}

We now derive the
scaling behavior at
the critical point,
where
the mass of the scalar field vanishes.
Then the coupling constants
must be
defined at
nonzero external momenta of the vertex functions.
For
$g$ we choose
 the normalization condition
\begin{equation}
\label{condg}
\Gamma^{(4)}_{1111}({\bf p}_1,{\bf p}_2,{\bf p}_3,{\bf p}_4)
|_{SP}=3\mu g,
\end{equation}
where the symbol $SP$ stands for the symmetry point\cite{HKSF}
\begin{equation}
{\bf p}_i\cdot{\bf p}_j=\frac{\mu^2}{4}(4\delta_{ij}-1).
\end{equation}
We shall distinguish the RG functions of
the
massive from those of the massless scaling regime
 by adding a tilde over the latter.
The  $\beta$-functions
of the
 gauge couplings have the same form as before.
The anomalous dimension of the vector field
changes, however, being now
\begin{equation}
\label{gammaa2}
\tilde{\gamma}_a=\frac{Nf}{32}.
\end{equation}
Also the beta function $\beta_g$
is now different, since the one-loop
gauge field graph with four external legs is now nonzero. This
leads to
an $f^2$-term in $\beta_g$:
\begin{equation}
\label{betag1}
\tilde{\beta}_{g}\equiv\mu\frac{\partial g}{\partial\mu}
=(2\tilde{\gamma}_\phi-1)g+\frac{N+8}{16}g^2+
\frac{\delta}{4\pi}f^2,
\end{equation}
where
\begin{widetext}
\begin{equation}
\label{gammaphi2}
\tilde{\gamma}_\phi=-\frac{f}{4\pi}\left[\frac{3\pi}{4t^2}+\frac{\pi}{2}-
\frac{3\pi t^2}{4}+3|t|-\frac{3}{|t|}
-\left(\frac{3}{2t^2}-1+\frac{3t^2}{2}\right)\arctan
\left(\frac{1-t^2}{2|t|}\right)\right],
\end{equation}
\begin{equation}
\label{omega}
\delta=\frac{\pi}{2t^2}+\frac{1}{|t|}-\frac{5\pi}{4}+
\left(-\frac{3}{2t^4}-\frac{4}{t^2}+8\right)\arctan\left(\frac{1}{2|t|}
\right)
+\left(\frac{3}{2t^4}+\frac{3}{t^2}-\frac{5}{2}\right)
\arctan\left(\frac{1-t^2}{2|t|}\right).
\end{equation}
In the limit $t\to 0$ we have 
$\tilde{\gamma}_\phi\to-f/4$ and $\delta\to 3\pi/2$, 
corresponding to the GL limit.

In contrast with the massive scaling regime, the present
equations yield a
charged
fixed point only for a limited range of $N$. The
beta functions vanish at
\begin{equation}
f_*=32/N,\,\label{g*1}
~~~~~~~~~~~~~~~~~g_{\pm}^*=\frac{8}{N\!+\!8}\left[1\!-\!2\tilde{\eta}\pm
\sqrt{(1\!-\!2\tilde{\eta})^2-\frac{160\delta_*}{\pi}}
\right],
\end{equation}
\end{widetext}
where $\tilde{\eta}\equiv\tilde{\gamma}_\phi(f_*,t_*)$ is
the anomalous dimension of the complex field
in the massless scaling regime. 
Remarkably,
there
 we find a tricritical
fixed point $g_-^*$, which is absent in the massive regime.
The two regimes are similar
for $\delta_*=0$, in which case there will be no tricritical
fixed point in both regimes.
This happens for $t_*=t_0\simeq 0.802693$. For
$t_*>t_0$ we find
$\delta_*<0$, in which case the tricritical point
becomes unstable, since it corresponds to $g_-^*<0$.
In Fig. \ref{Figomega} we plot $\delta$ as a function of $t$.

The charged fixed
points are accessible only if $t_*\geq \tilde{t}_c(N)$, where
$\tilde{t}_c(N)$ is the value
of $t_*$ that vanishes the discriminant in Eq. (\ref{g*1}).
For example, if $N=10$ we have $\tilde{t}_c(10)\simeq 0.752751$.
>From Eq.~(\ref{tineq}) we find $t_c(10)\simeq 0.788854$, and
therefore $\tilde{t}_c(10)<t_c(10)$.

In order to calculate the $\tilde{\nu}$-exponent, we need the RG function
$\tilde{\gamma}_\phi^{(2)}$. This function is much more complicate in
the massless scaling regime and is given explicitly by

\begin{figure}
\begin{picture}(87.7,143.05)
\unitlength.7mm
\put(-36,3){\includegraphics{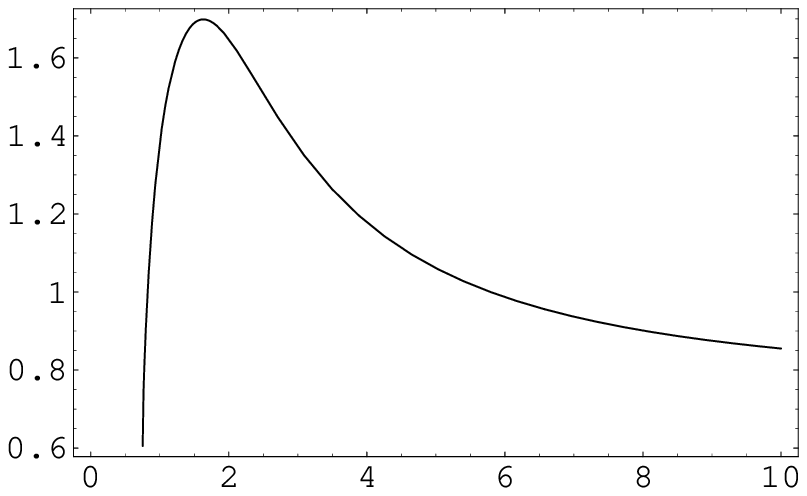}}
\put(26,0.5){$t_*$}
\put(-40,40){$\tilde{\nu}$}
\end{picture}
\caption{\label{Fignu2}Plot of the critical exponent $\tilde{\nu}$
as a function of $t_*$ for $N=10$.}
\end{figure}

\begin{widetext}
\begin{equation}
\label{gamma22}
\tilde{\gamma}_\phi^{(2)}=-\frac{(N+2)g}{16}
-\frac{f}{4\pi}
\left\{
\pi\frac{4t^2-3}{4\sqrt{3}t^2}
+\frac{4t^2+3}{2\sqrt{3}t^2}\arctan\left(\frac{3-4t^2}{4\sqrt{3}|t|}
\right)
+\frac{(3-4t^2)(3+4t^2)}{8|t|\Delta}\left[1+
\frac{|t|}{\sqrt{\Delta}}\arctan\left(\frac{\sqrt{\Delta}}{|t|}\right)
\right]\right\},
\end{equation}
\end{widetext}
where
\begin{equation}
\Delta\equiv t^4+\frac{t^2}{2}+\frac{9}{16}.
\end{equation}
In Fig. \ref{Fignu2}
we plot $\tilde{\nu}$ as a function of $t_*\geq t_c(10)$. The curve
has a maximum for $t_*\simeq 1.631$, where
 $\tilde{\nu}_{\rm max}=1.7$.

\section{
Sign of anomalous dimension $\eta$ and large-$N$ limit
}

A much  debated topic in the GL theory
is the
physical meaning of
a negative sign of the $\eta$-exponent found in
analytic calculations\cite{deCalan1,Nogueira,KleinNog1} and computer simulations
\cite{Hove}. In early discussions of the subject it
had been argued
that a negative  $\eta$ would be unphysical
since it would
violate the K\"allen-Lehmann spectral
representation \cite{Kiometzis1,deCalan1}.
However, it is now being understood
that
a negative sign of $\eta$ in the GL model
makes sense.
In fact, there
are several  physical systems
where negative $ \eta $s have been found before,
most prominently
  magnetic systems,
which
show
strong momentum space instabilities \cite{Selke}. These
can produce
 a non-uniform phase with
a modulated order parameter. The point where
the
modulated phase sets in is called a
Lifshitz point \cite{Hornreich}. Physical systems with
a Lifshitz point
have a negative $\eta$-exponent.
It has recently
been  argued
by us \cite{Nogueira,KleinNog1} that
such
momentum space instabilities
occur also
in superconductors,
 implying the existence of a Lifshitz point in
the phase diagram,
and thus explaining
 the negative sign
of $\eta$. A different explanation has been given recently in 
Ref. \onlinecite{Mo}, focusing on the geometric properties of 
the critical fluctuations. There the anomalous dimension 
is related to the Hausdorff dimension of the critical 
fluctuations \cite{Mo}.

The CS term in the GL model is expected to  affect this picture, since for
infinite $t_*$, the gauge field decouples from the scalar field. In this
limit, the critical exponents are those of a
pure scalar field
theory which has $\eta>0$. Thus we may wonder at which
finite
value of $t_*$ the sign change of  $\eta$ occurs.

\begin{figure}
\begin{picture}(87.87,143.05)
\unitlength.7mm
\put(-36,3){\includegraphics{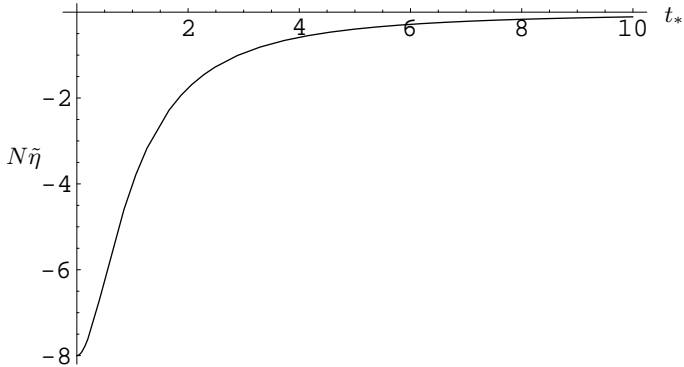}}
\put(83,70){$t_*$}
\put(-42,43){$N\tilde{\eta}$}
\end{picture}
\caption{\label{Figeta2}Plot of $N\tilde{\eta}$ as a function of $t_*$.}
\end{figure}

In Fig. \ref{Figeta2} we plot
$N\tilde{\eta}$ as a function of $t_*$. We see that at
to one-loop  order,
$\tilde{\eta}$ is always negative and approaches zero
for $t_*\rightarrow \infty$.  It would be desirable to know
the
two-loop corrections
in the
massless scaling regime and check if there exists a finite
value of $t_*$ where the sign of $\tilde{\eta}$ changes.
We
 have not yet done this
calculation due to its complexity
for arbitrary
 $t$.
We can, however, easily write down  $\eta$ in the
limit of large $N$
for all coupling strengths
in the
massive scaling regime.
Then the
 CSGL model is in the same
 universality
class as
the $CP^{N/2-1}$ model with a CS term,
and here the critical exponents
have been computed by Rajeev and Ferretti \cite{Ferretti}.
The result for $\eta$ is
\begin{equation}
\label{CPCS}
\eta=-\frac{40}{\pi^2 N}\left(1-\frac{16}{15}\frac{\bar{t}^2}{
1+\bar{t}^2}\right),
\end{equation}
where $\bar{t}=4t/\pi$. In the limit $\bar{t}\to\infty$
this reduces correctly
to
 $\eta$ of the $O(N)$-symmetric scalar model\cite{HKSF,ZJ}
to  order
$1/N$.

\begin{figure}
\begin{picture}(87.87,143.05)
\unitlength.7mm
\put(-36,3){\includegraphics{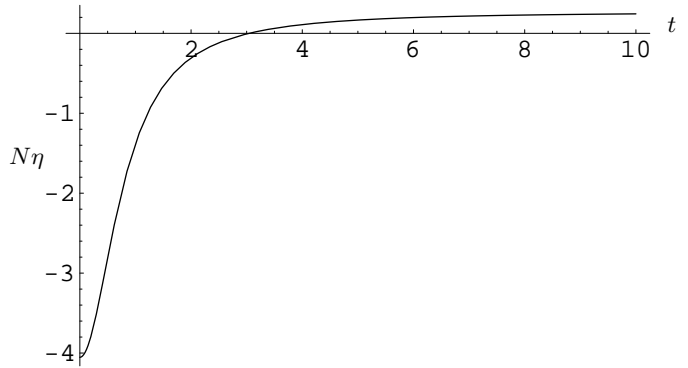}}
\put(83,68){$t$}
\put(-42,43){$N\eta$}
\end{picture}
\caption{\label{Figeta3}Plot of $N\eta$, where $\eta$ is given
by Eq. (\ref{CPCS}), as a function of $t$.}
\end{figure}

{}From Eq. (\ref{CPCS}) we see that $\eta$ changes sign at
$\bar{t}=\sqrt{15}$. In Fig. \ref{Figeta3} we plot $N\eta$ as given in
Eq. (\ref{CPCS}) as a function of $t$. Interestingly,
Figs. \ref{Figeta2} and \ref{Figeta3}
 look very similar for low $t$,
 up to a factor of two
in the vertical scale, although
the two curves come from two completely different
 approximations.  In addition, the comparison teaches us that
 the sign of $\eta$ may easily change
in perturbation theory ny including
 higher-order corrections.

\begin{figure}
\begin{picture}(87.87,143.05)
\unitlength.7mm
\put(-36,3){\includegraphics{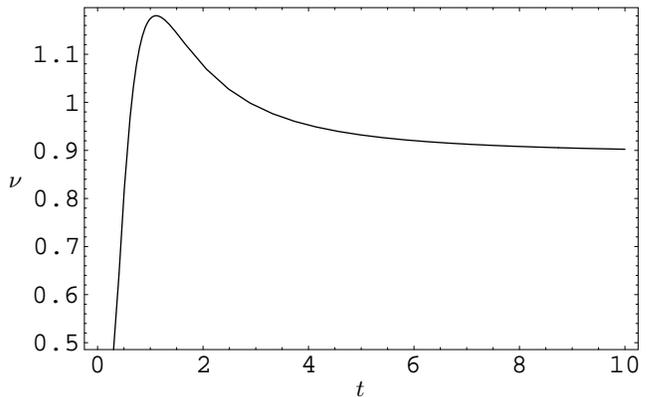}}
\put(26,0.5){$t$}
\put(-40,40){$\nu$}
\end{picture}
\caption{\label{Fignu3}Plot of $\nu$, where $\nu$ is given
by Eq. (\ref{CPCSnu}) with $N=10$, as a function of $t$.}
\end{figure}

It is also interesting to consider the critical exponent $\nu$
to leading order in
$1/N$. We have
\begin{equation}
\label{CPCSnu}
\nu=1-\frac{96}{\pi^2N}\left[1-\frac{8}{9}\frac{\bar{t}^2(\bar{t}^2+4)}{
(1+\bar{t}^2)^2}\right].
\end{equation}
We have plotted the $\nu$-exponent given above in Fig. \ref{Fignu3}
for $N=10$, for the sake of comparison with
the one-loop result
for
the massless scaling regime. There is again
a remarkable
similarity between this curve
and the one
for $\tilde{\nu}$ in Fig. \ref{Fignu2}. This striking resemblance between
the massive scaling regime
at order $1/N$ and
the massless scaling regime at one-loop seems to indicate that
perturbation theory within
the massive scaling regime is worse than perturbation
theory within
the
massless scaling regime, and that the exact curves
in the two regimes
 may ultimately coincide.

\section{The ``fermionic'' fixed point}

Let us see what we can learn from our scaling study about
 strongly interacting fermions. General arguments
involving bosonization \cite{Fradkin} and duality transformations
\cite{KleinNog} indicate that the fixed points corresponding to
fermions lie, in
the
 massless scaling regime, at $t_*=1/2\pi$ and $f_*=32/N$. Then there
are no fixed points $g_\pm^*$ for the physical case  $N=2$. We know, however,
from duality arguments applied to the GL model
\cite{KleinertBook,Dasgupta,Kleinert} that this
one-loop result is not trustworthy
 for $N=2$. Within
the
 massless scaling regime it is possible to
obtain a charged fixed point for all values of $N$ at the one-loop
level by
introducing a new arbitrary parameter $c$ which corresponds to
the ratio between the two renormalization scales defining the
gauge and scalar couplings, respectively\cite{Herbut}.
This procedure was also applied to the CSGL model
in Ref. \onlinecite{deCalan}. A charged fixed point
 is found
 for $N=2$
if $c$ is chosen large enough. This happens since
 $\gamma_a$
is modified
to
\begin{equation}
\gamma_a=\frac{cNf}{32}.
\end{equation}
Since $ \gamma _a=1$ at the fixed point,\mn{check this sentence}
a large
$c$ makes $f_*$ sufficiently small
 to reach the
fixed point for $N=2$. The main drawback of this technique is the
fact that $c$ is not determined by the formalism.
In Ref.~\onlinecite{Herbut} it was chosen to reproduce the tricritical
point determined in Ref.~\onlinecite{Kleinert} by a disorder field theory
of the GL model.

Recently, we have succeeded in obtaining a charged fixed point at
$N=2$ by defining a new RG approach in the ordered phase
\cite{KleinNog1}, where the two
length scales of the GL model are well defined by
the correlation length $\xi$ and
the penetration depth $\lambda$. This makes the parameter
$c$ in  Refs. \onlinecite{Herbut} and
\onlinecite{deCalan} superfluous.
The application of Sothis calculation procedure
to the
Lagrangian (\ref{Lspin})
is complicated due to the
CS term. This creates a gauge
field propagator in the ordered phase
with
two
different masses, the
CS mass and another one generated by the Higgs mechanism.
To avoid this complication we
shall restrict ourselves here
to the $c$-approach.
The constant $c$ will be fixed
 by demanding that
in the $\theta=0$ -model
the critical exponent $\nu$ has a $XY$ value, as found
in the duality approach \cite{Kiometzis}. For the
 $XY$ value $ \nu \simeq0.67$, this fixes
$c\simeq 82.7$. In Ref.~\onlinecite{deCalan}, a smaller value of
$c$ was used by approximating the RG function $\tilde{\nu}_\phi$
 by
$\tilde{\nu}_\phi\simeq
(1-\tilde{\gamma}_\phi^{(2)}/2+\tilde{\gamma}_\phi/2)/2$.
However, this approximation gives $\tilde{\nu}=0.6$ for
$f_*=t_*=0$, while if we don't use such an approximation we obtain
a much better value in this limit, $\tilde{\nu}=0.625$, which is
just the one-loop value for the $O(2)$-symmetric $\phi^4$ theory\cite{HKSF}.

Therefore we obtain, with $c=82.7$ and $t_*=1/2/\pi$:
\begin{equation}
\tilde{\eta}\simeq -0.05,
\end{equation}
\begin{equation}
\tilde{\nu}\simeq 0.66.
\end{equation}
We see that the ``fermionic'' fixed point is not much different
from the GL fixed point
for the value
of $c$ under consideration.

\section{Conclusion}

We have studied and compared two scaling
regimes in the CSGL model.
In the
massive scaling regime, charged fixed points exist for all values of $N$.
However, not all of them  lead to physical values of the
critical exponents which restrict the range of allowed
values of $N$. By restricting the values of $t_*$ we have
been able to obtain physical exponents to all $N$ in
the
massive scaling regime.
The interval of admissible $t_*$ is obtained from the
inequality (\ref{tineq}).

In
the
 massless scaling regime we find a similar
restriction through a more involved inequality, since
$\tilde{\eta}$ has a far more complicate expression.

The discussion in the
massive scaling regime
yields no tricritical point, a result consistent with
the Landau expansion of the mean-field free energy in
Eq. (\ref{expHLMCS}). However, from the non-expanded mean-field
free energy (\ref{HLMCS}) it is seen that for sufficiently
small $\theta$ we obtain a  first-order phase transition.
In particular,
we recover the usual HLM result for $\theta=0$,
in contrast to  Eq. (\ref{expHLMCS}) which does not
have the correct $\theta\rightarrow 0$ -limit.

The
 behavior in the massless scaling regime
is more consistent with Eq. (\ref{HLMCS})
since it exhibits a tricritical fixed point for $t_*<t_0$. In the
region $t_*\geq t_0$, the two scaling regimes
 look quite similar, at
least qualitatively. The $1/N$-expansion applied to
the
massive scaling regime makes this similarity even greater and suggest
that perturbation theory applied in
the
 massless scaling regime is better behaved than in the
massive regime.

An interesting point with respect to the $1/N$-expansion in
the
massive scaling regime
is the sign change in $\eta$ for $\bar{t}=\sqrt{15}$. This
never happens for a GL model \cite{Hove,Nogueira,KleinNog1}. Inspired
by our recent work suggesting that the sign of $\eta$ is related
to momentum space instabilities \cite{Nogueira,KleinNog1}, we may
conjecture that when an external magnetic field is included in the
CSGL model,  vortex lattices should
not exist  above a certain critical value
of the topological mass.

We have discussed briefly what we called ``fermionic'' fixed point,
that is, the fixed point where the CSGL model corresponds
to bosonized three-dimensional interacting fermions.
At this fixed point $t_*=1/2\pi$.
Unfortunately, $g_\pm^*$ is not real in this case
if $N=2$. In order to reach a charged fixed point for $N=2$
we introduced an arbitrary parameter $c$ corresponding
to the ratio between the renormalization points of the gauge
couplings and scalar coupling. The value of $c$ has been fixed
in the $t=0$ model. As $t$ is turned on to $t_*=1/2\pi$ the
values of the critical exponents doesn't show an appreciable
change with respect to the $t=0$ case.

\begin{acknowledgments}
The work of FSN is supported by the Alexander von Humboldt foundation.
\end{acknowledgments}

\end{document}